\documentclass[11pt]{article}

\usepackage{cite,color,xspace}
\usepackage{latexsym}
\usepackage{todonotes}
\usepackage{amsmath}
\usepackage{amsfonts}

\usepackage{tikz,tikz-qtree,tikz-qtree-compat}

\usepackage{amsthm}
\usepackage{cite}
\usepackage{url}
\usepackage{graphicx}
\usepackage{subfigure}
\usepackage{algorithm}
\usepackage[noend]{algpseudocode}
\usepackage[matha]{mathabx}
\usepackage{txfonts} % for semijoin                                             

\newcommand{\dlm}{\textsf{DLM}\xspace}
\newcommand{\agm}{\textsf{AGM}\xspace}
\newcommand{\nprr}{\textsf{NPRR}\xspace}
\newcommand{\leapfrog}{\textsf{Leapfrog Triejoin}\xspace}
\newcommand{\glvv}{\textsf{GLVV}\xspace}

\newcommand{\atoms}{\text{atoms}}

\newcommand{\calH}{\mathcal H}
\newcommand{\calD}{\mathcal D}

\newcommand{\calV}{\mathcal V}
\newcommand{\calX}{\mathcal X}
\newcommand{\calY}{\mathcal Y}
\newcommand{\calE}{\mathcal E}

\newcommand{\eat}[1]{}

\renewcommand{\Pr}{\textnormal{P}}

\newcommand{\be}{\begin{enumerate}}
\newcommand{\ee}{\end{enumerate}}
\newcommand{\bi}{\begin{itemize}}
\newcommand{\ei}{\end{itemize}}
\newcommand{\beq}{\begin{equation}}
\newcommand{\eeq}{\end{equation}}

\newcommand{\bp}{\begin{proof}}
\newcommand{\ep}{\end{proof}}
\newcommand{\bcor}{\begin{cor}}
\newcommand{\ecor}{\end{cor}}
\newcommand{\bthm}{\begin{thm}}
\newcommand{\ethm}{\end{thm}}
\newcommand{\blmm}{\begin{lmm}}
\newcommand{\elmm}{\end{lmm}}
\newcommand{\bdefn}{\begin{defn}}
\newcommand{\edefn}{\end{defn}}
\newcommand{\bprop}{\begin{prop}}
\newcommand{\eprop}{\end{prop}}
\newcommand{\bconj}{\begin{conj}}
\newcommand{\econj}{\end{conj}}
\newcommand{\bopm}{\begin{opm}}
\newcommand{\eopm}{\end{opm}}
\newcommand{\brmk}{\begin{rmk}}
\newcommand{\ermk}{\end{rmk}}

\newcommand{\suchthat}{\ | \ }

\newcommand{\vars}{\textnormal{vars}}

\newcommand{\mv}[1]{\mathbf{#1}}

\theoremstyle{plain}                   % default
\newtheorem{thm}{Theorem}[section]
\newtheorem{lmm}[thm]{Lemma}
\newtheorem{prop}[thm]{Proposition}
\newtheorem{cor}[thm]{Corollary}

\theoremstyle{definition}              % Examples and all
\newtheorem{example}{Example}

\newtheorem{opm}[thm]{Open Problem}
\newtheorem{conj}[thm]{Conjecture}

\newtheorem{defn}[thm]{Definition}

\newtheorem{rmk}[thm]{Remark}

\newcommand{\bbox}{
\begin{center}
\begin{tabular}{|c|}
\hline
}
\newcommand{\ebox}{
\\
\hline
\end{tabular}
\end{center}
}

\newlength{\toppush}
\algrenewcommand\algorithmicrequire{\textbf{Input:}}
\algrenewcommand\algorithmicensure{\textbf{Output:}}
\algrenewcommand\algorithmicwhile{\textbf{While}}
\algrenewcommand\algorithmicfor{\textbf{For}}
\algrenewcommand\algorithmicreturn{\textbf{Return}}
\algrenewcommand\algorithmicif{\textbf{If}}

\newcommand{\ms}{\text{Minesweeper}\xspace}

\newcommand{\D}{\mathbf{D}}

\usepackage[margin=1in]{geometry}

\begin{document}

\pagestyle{empty}

\title{Skew Strikes Back: New Developments in the Theory of Join Algorithms}

\author{Hung Q. Ngo\\
University at Buffalo, SUNY\\
{hungngo@buffalo.edu} 
\and
Christopher R\'{e}\\
Stanford University\\
{chrismre@cs.stanford.edu}
\and
Atri Rudra\\
University at Buffalo, SUNY\\
{atri@buffalo.edu}
}

%\date{}

\maketitle

%% \begin{abstract}
%% We survey some recent development in join algorithms.
%% \end{abstract}

% A category with the (minimum) three required fields
%\category{H.4}{Information Systems Applications}{Miscellaneous}
%A category including the fourth, optional field follows...
%\category{D.2.8}{Software Engineering}{Metrics}[complexity measures, 
%performance measures]

%\terms{Delphi theory}

%\keywords{ACM proceedings, \LaTeX, text tagging}

% ----------------------------------------------------------------------------
\section{Introduction}
% ----------------------------------------------------------------------------

Evaluating the relational join is one of the central algorithmic and
most well-studied problems in database systems. A staggering number of
variants have been considered including Block-Nested loop join,
Hash-Join, Grace, Sort-merge (see Grafe~\cite{graefe93} for a survey,
and \cite{Blanas:2011:DEM:1989323.1989328,
  Kim:2009:SVH:1687553.1687564, Chaudhuri:1998:OQO:275487.275492} for
discussions of more modern issues).  Commercial database engines use
finely tuned join heuristics that take into account a wide variety of
factors including the selectivity of various predicates, memory, IO,
etc. In spite of this study of join queries, the textbook description
of join processing is {\em suboptimal}. This survey describes recent
results on join algorithms that have provable worst-case optimality
runtime guarantees.  We survey recent work and provide a simpler and
unified description of these algorithms that we hope is useful for
theory-minded readers, algorithm designers, and systems implementors.

Much of this progress can be understood by thinking about a simple
join evaluation problem that we illustrate with the so-called
{\em triangle query}, a query that has become increasingly popular in the 
last decade with the advent of social networks, biological motifs, and 
graph databases~\cite{DBLP:conf/www/SuriV11,
Tsourakakis:2008:FCT:1510528.1511415}

\begin{quote}
{\it Suppose that one is given a graph with $N$ edges, how many distinct
triangles can there be in the graph?}
\end{quote}

\noindent
A first bound is to say that there are at most $N$ edges, and hence at
most $O(N^{3})$ triangles. A bit more thought suggests that every
triangle is indexed by any two of its sides and hence there at most
$O(N^{2})$ triangles. However, the correct, tight, and non-trivial
asymptotic is $O(N^{3/2})$. An example of the questions this survey is
how do we list all the triangles in time $O(N^{3/2})$?  Such an
algorithm would have a worst-case optimal running time.  In contrast,
traditional databases evaluate joins pairwise, and as has been noted
by several authors, this forces them to run in time $\Omega(N^2)$ on
some instance of the triangle query. This survey gives an overview of
recent developments that establish such non-trivial bounds for {\em
  all} join queries and algorithms that meet these bounds, which we
call worst-case optimal join algorithms.

Estimates on the output size of join have been known since the 1990s,
thanks to the work of Friedgut and Kahn \cite{MR1639767} in the
context of bounding the number of occurrences of a given small
hypergraph inside a large hypergraph. More recently and more
generally, tight estimates for the natural join problem were derived
by Grohe-Marx~\cite{GM06} and Atserias-Grohe-Marx~\cite{AGM08} (\agm
henceforth).  In fact, similar bounds can be traced back to the 1940s
in geometry, where it was known as the famous Loomis-Whitney
inequality~\cite{MR0031538}.  The most general geometric bound is by
Bollob\'as-Thomason in the 1990s~\cite{MR1338683}. We proved (with
Porat) that \agm and the discrete version of Bollob\'as-Thomason are {\em
  equivalent}~\cite{NPRR}, and so the connection between these areas
is deep.

Connections of join size to arcane geometric bounds may reasonably
lead a practitioner to believe that the cause of suboptimality is a
mysterious force wholly unknown to them---but it is not; it is the old
enemy of the database optimizer, skew. We hope to highlight two conceptual
messages with this survey:

\begin{itemize}

\item The main ideas of the the algorithms presented here are an
  optimal way of avoiding skew -- something database practitioners
  have been fighting with for decades. We describe a theoretical basis
  for one family of techniques to cope with skew by relating them to
  geometry.

\item The second idea is a challenge to the database dogma of doing
  {\it ``one join at a time,''} as is done in traditional database
  systems. We show that there are classes of queries for which
  {\em any} join-project plan is destined to be
  slower than the best possible run time by a polynomial factor {\em
    in the data size}.
\end{itemize}

\noindent
\emph{Outline of the Survey.}  We begin with a short (and necessarily
incomplete) history of join processing with a focus on recent
history. In Section~\ref{sec:triangle}, we describe how these
algorithms work for the triangle query. In Section~\ref{sec:prelim},
we describe how to use these new size bounds for join queries. In
Section~\ref{sec:agm-prf-algo}, we provide new simplified proofs of
these bounds and join algorithms. Finally, we describe two open
questions in Section~\ref{sec:concl}

\subsection*{A Brief History of Join Processing} 
Conjunctive query evaluation in general and join query evaluation in particular
%is a fundamental problem in database theory, which
 have a very long history and deep connections to logic and constraint
 satisfaction
 \cite{DBLP:conf/stoc/ChandraM77,DBLP:journals/jacm/Fagin83,
   DBLP:conf/stoc/Vardi82,DBLP:journals/jcss/GottlobLS02,
   DBLP:journals/jacm/GottlobMS09,DBLP:journals/tcs/ChekuriR00,
   DBLP:journals/jcss/KolaitisV00,DBLP:conf/pods/PapadimitriouY97}.
 Most of the join algorithms with provable performance guarantees work
 for specific classes of queries.\footnote{Throughout this survey, we
   will measure the run time of join algorithms in terms of the input
   data, assuming the input query has constant size; this is known as
   the {\em data complexity} measure, which is standard in database
   theory \cite{DBLP:conf/stoc/Vardi82}.} As we describe, there are
 two major approaches for join processing: using {\em structural
   information of the query} and {\em using cardinality
   information}. As we explain, the \agm bounds are exciting because
 they bring together both types of information.

\paragraph*{The Structural Approaches} 
On the theoretical side, many algorithms use some structural property
of the query such as {\em acyclicity} or {\em bounded ``width.''}  For
example, when the query is acyclic, the classic algorithm of
Yannakakis~\cite{DBLP:conf/vldb/Yannakakis81} runs in time essentially
linear in the input plus output size.  A query is acyclic if and only
if it has a {\em join tree}, which can be constructed using the
textbook {\em GYO-reduction} \cite{yu1984determining,
  ontheuniversalrelation}
%, Tarjan:1984:SLA:1169.1179}.
%, DBLP:conf/pods/PapadimitriouY97, }
%DBLP:conf/focs/ChandraH80,DBLP:conf/stoc/ChandraH79,
%DBLP:conf/pods/Vardi95}.

Subsequent work further expand the classes of queries which can be
evaluated in polynomial time.  These work define progressively more
general notions of ``width'' for a query, which intuitively measures
how far a query is from being acyclic. Roughly, these results state
that if the corresponding notion of ``width'' is bounded by a
constant, then the query is ``tractable,'' i.e. there is a polynomial
time algorithm to evaluate it.  For example, Gyssens et
al. \cite{DBLP:journals/ai/GyssensJC94, DBLP:conf/adbt/GyssensP82}
showed that queries with bounded ``degree of acyclicity'' are
tractable.  Then come {\em query width} (qw) from Chekuri and
Rajaraman~\cite{DBLP:journals/tcs/ChekuriR00}, {\em hypertree width}
and {\em generalized hypertree width} (ghw) from Gottlob et al.
\cite{DBLP:journals/sigmod/Scarcello05,
  DBLP:journals/jcss/GottlobLS03}.  These are related to the {\em
  treewidth} (tw) of a query's hypergraph, rooted in Robertson and
Seymour on graph minors \cite{MR855559}.  Acyclic queries are exactly
those with qw $=1$.

\paragraph*{Cardinality-based Approaches } 
Width only tells half of the story, as was wonderfully articulated in
Scarcello's SIGMOD Record paper
\cite{DBLP:journals/sigmod/Scarcello05}:

\begin{quote} {\em decomposition methods
  focus ``only'' on structural features, while they completely
  disregard ``quantitative'' aspects of the query, that may
  dramatically affect the query-evaluation time.}  
\end{quote}

\noindent
Said another way, the width approach disregards the input relation
sizes and summarizes them in a single number, $N$. As a
result, the run time of these structural approaches is $O(N^{w+1}\log
N)$, where $N$ is the input size and $w$ is the corresponding width
measure. On the other hand, commercial RDBMSs seem to place little
emphasis on the structural property of the query and tremendous
emphasis on the cardinality side of join processing. Commercial
databases often process a join query by breaking a complex multiway
join into a series of pairwise joins; an approach first described in
the seminal System R, Selinger-style optimizer from the
1970~\cite{Selinger:1979:APS:582095.582099}. However, throwing away
this structural information comes at a cost: {\em any} join-project
plan is destined to be slower than the best possible run time by a
polynomial factor {\em in the data size}. 

\paragraph*{Bridging This Gap} 
A major recent result from \agm \cite{AGM08, GM06} is the key to
bridging this gap: \agm derived a {\em tight} bound on the output size
of a join query as a function of individual input relation sizes {\em
  and} a much finer notion of ``width''.  The \agm bound leads to the
notion of {\em fractional query number} and eventually {\em fractional
  hypertree width} (fhw) which is strictly more general than all of
the above width notions \cite{Marx:2010:AFH:1721837.1721845}.  To
summarize, for the same query, it can be shown that
\[ \text{fhw} \leq \text{ghw} \leq \text{qw} \leq \text{tw} + 1, \]
%there are queries for which these inequalities are strict,
and the join-project algorithm from \agm runs in time
$O(N^{\text{fhw}+1}\log N)$. %More interestingly, these bounds
\agm's bound is sharp enough to take into account cardinality information, and
they can be {\em much} better when the input relation sizes vary. 
The bound takes into account {\em both} the input relation
statistics {\em and} the structural properties of the query. The
question is whether it is possible and how to turn the bound into
join algorithms, with runtime $O(N^{\text{fwh}})$ and much better
when input relations do not have the same size.

The first such worst-case optimal join algorithm was designed by the authors 
(and Porat) in 2012~\cite{NPRR}. Soon after, an algorithm (with a simpler
description) with a similar optimality guarantee was presented soon
after called ``\leapfrog''~\cite{leapfrog}. Remarkably this
algorithm was implemented in a commercial database system {\em before}
its optimality guarantees were discovered. A key idea in the
algorithms is handling skew in a theoretically optimal way, and uses
many of the same techniques that database management systems have used
for decades heuristically~\cite{Xu:2008:HDS:1376616.1376720,DeWitt:1992:PSH:645918.672512,Walton:1991:TPM:645917.672307}

A technical contribution of this survey is to describe the algorithms
from \cite{NPRR} and \cite{leapfrog} and their analyses in one
unifying (and simplified) framework.  In particular, we make the
observation that these join algorithms are in fact special cases of a
{\em single} join algorithm.  This result is new and serves to explain
the common link between these join algorithms. We also illustrate some
unexpected connections with geometry, which we believe are interesting
in their own right and may be the basis for further theoretical
development.

% ----------------------------------------------------------------------------
\section{Much ado about the triangle}
\label{sec:triangle}
% ----------------------------------------------------------------------------

We begin with the triangle query
\[Q_{\triangle}=R(A,B)\Join S(B,C)\Join T(A,C).\]
The above query is the simplest cyclic query and is rich enough to
illustrate most of the ideas in the new join algorithms.\footnote{This
query can be used to list all triangles in a given graph $G=(V,E)$,
if we set $R,S$ and $T$ to consist of all pairs $(u,v)$ and $(v,u)$
for which $uv$ is an edge. Due to symmetry, each triangle in $G$ will be 
listed $6$ times in the join.} 
We first describe the traditional way to evaluate this query and how skew 
impacts this query. We then develop two closely related algorithmic ideas 
allowing us to mitigate the impact of skew in these examples; they are the 
key ideas behind the recent join processing algorithms.

\subsection{Why traditional join plans are suboptimal} 

The textbook way to evaluate any join query, including
$Q_{\triangle}$, is to determine the best pair-wise join
plan~\cite[Ch.~15]{cowbook}. Figure~\ref{fig:example} illustrates three
plans that a conventional RDBMS would use for this query. 
For example, the first plan is to compute the intermediate join 
$P=R\Join T$ and then compute $P\Join S$ as the final output. 

\begin{figure}[h!]
\begin{center}
\begin{tikzpicture}
\Tree[.{$\Join$} [.{$\Join$} {$\textcolor{red}{R}$} {$\textcolor{blue}{T}$} ] {$\textcolor{green}{S}$} ]
\end{tikzpicture}
\hspace*{7mm}
\begin{tikzpicture}
\Tree[.{$\Join$} [.{$\Join$} {$\textcolor{red}{R}$} {$\textcolor{green}{S}$} ] {$\textcolor{blue}{T}$} ]
\end{tikzpicture}
\hspace*{7mm}
\begin{tikzpicture}
\Tree[.{$\Join$} [.{$\Join$} {$\textcolor{green}{S}$} {$\textcolor{blue}{T}$} ] {$\textcolor{red}{R}$} ]
\end{tikzpicture}
\end{center}
\caption{The three pair-wise join plans for $Q_{\triangle}$.}
\label{fig:example}
\end{figure}

\begin{figure*}[t]
\begin{center}
\begin{minipage}{.5\textwidth}
  \centering
\[R=\{a_0\}\times \{b_0,\dots,b_m\}\cup \{a_0,\dots,a_m\}\times \{b_0\}\]
\[S=\{b_0\}\times \{c_0,\dots,c_m\}\cup \{b_0,\dots,b_m\}\times \{c_0\}\]
\[T=\{a_0\}\times \{c_0,\dots,c_m\}\cup \{a_0,\dots,a_m\}\times \{c_0\}\]
%\caption{Bad Example}
%\label{fig:bad-eg}
\end{minipage}%
\begin{minipage}{.5\textwidth}
  \centering
 \scalebox{.2}{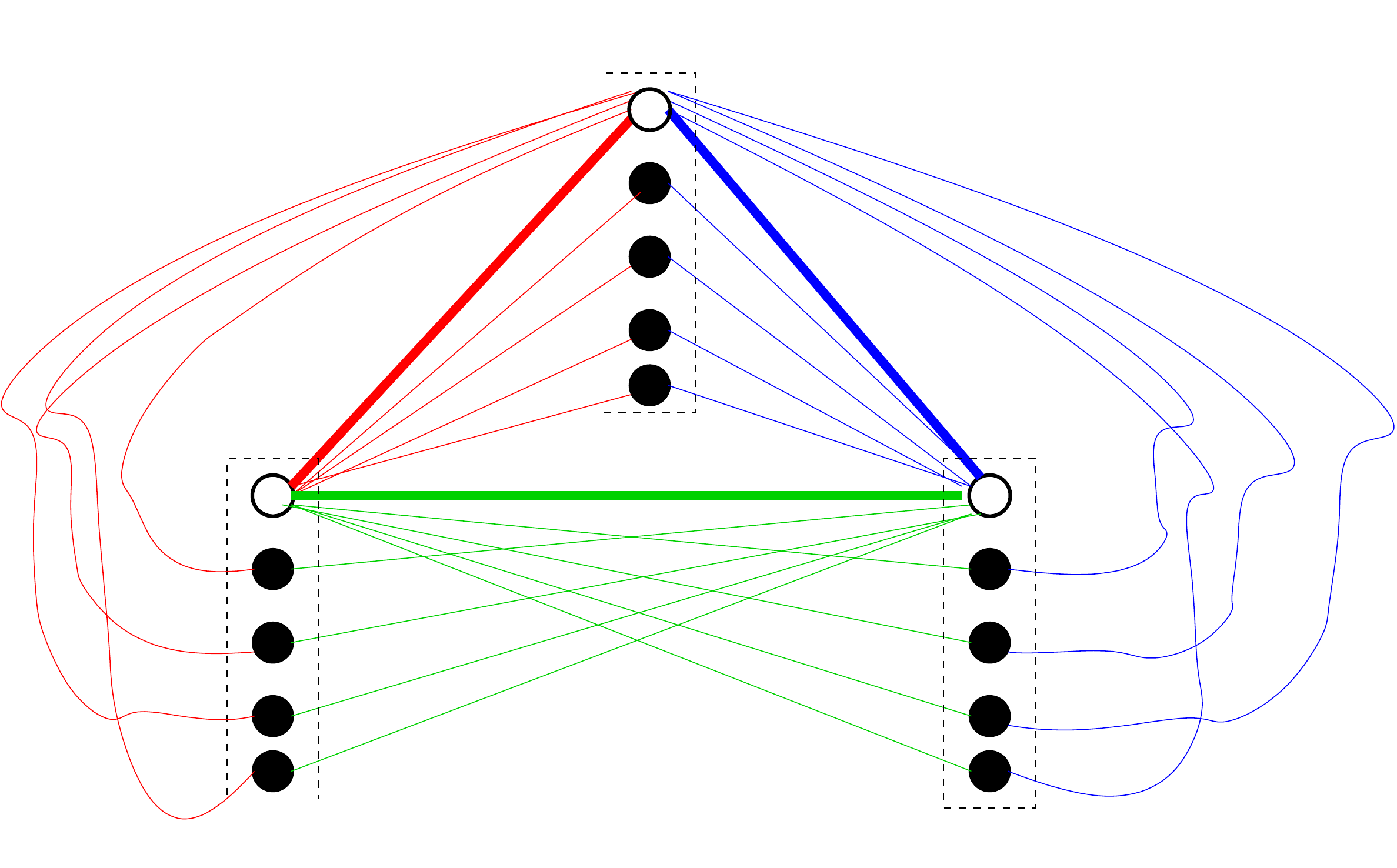}
%\caption{Illustration for bad example for $m=4$.}
%\label{fig:triangle-bad}
\end{minipage}
\caption{Counter-example for join-project only plans for the triangles (left) and an illustration for $m=4$ (right). The pairs connected by the red/green/blue edges form the tuples in the relations $R$/$S$/$T$ respectively. Note that the in this case each relation has $N=2m+1=9$ tuples and there are $3m+1=13$ output tuples in $Q_{\triangle}$. Any pair-wise join however has size $m^2+m=20$.}
\label{fig:triangle-bad}
\end{center}
\end{figure*}

We next give a family of instances for which any of the above three
join plans must run in time $\Omega(N^{2})$ because the intermediate relation
$P$ is too large.
Let $m\ge 1$ be a positive integer. 
The instance family is illustrated in Figure~\ref{fig:triangle-bad}, where
the domain of the attributes $A,B$ and $C$ are $\{a_0,a_1,\dots,a_m\}$,
$\{b_0,b_1,\dots,b_m\}$, and $\{c_0,c_1,\dots,c_m\}$ respectively. 
In Figure~\ref{fig:triangle-bad}, the unfilled circles denote the values
$a_0,b_0$ and $c_0$ respectively while the black circles denote the
rest of the values. 

For this instance each relation has $N=2m+1$ tuples and
$|Q_{\triangle}|=3m+1$; however, any pair-wise join has size $m^2+m$.
Thus, for large $m$, any of the three join plans will take 
$\Omega(N^2)$ time. In fact, it can be shown that even if we allow 
projections in addition to joins, the $\Omega(N^2)$ bound still holds. 
(See Lemma~\ref{LEM:BAD:INSTANCE}.)
By contrast, the two algorithms shown in the next section runs in time $O(N)$,
which is optimal because the output itself has $\Omega(N)$ tuples!

\subsection{Algorithm 1: The Power of Two Choices} 

Inspecting the bad example above, one can see a root cause for the
large intermediate relation: $a_0$ has ``high degree" or in the
terminology to follow it is {\em heavy}. In other words, it is an
example of {\em skew}.  To cope with skew, we shall take a strategy
often employed in database systems: we deal with nodes of high and low
skew using different join techniques~\cite{DeWitt:1992:PSH:645918.672512,Xu:2008:HDS:1376616.1376720}.  The first goal
then is to understand when a value has high skew.  To shorten
notations, for each $a_i$ define
\[ Q_\triangle[a_i] := \pi_{B,C} (\sigma_{A=a_i} (Q_{\triangle})). \]
We will call $a_i$ {\em heavy} if 
\[ |\sigma_{A=a_i}(R\Join T)|\ge |Q_\triangle[a_i]|. \] 
In other words, the value $a_i$ is {\em heavy} if its contribution to the size of
intermediate relation $R \Join S$ is {\em greater} than its contribution to 
the size of the output. 
Since 
\[ |\sigma_{A=a_i} (R\Join S)| = |\sigma_{A=a_i} R| \cdot |\sigma_{A=a_i} S|, \]
we can easily compute the left hand side of the above inequality from
an appropriate index of the input relations.
Of course, we do not know $|Q_\triangle[a_i]|$ until after we have computed 
$Q_\triangle$. However, note that we always have 
$Q_\triangle[a_i]\subseteq S$. Thus, we will use $|S|$
as a proxy for $|Q_\triangle[a_i]|$. The two choices come from the following 
two ways of computing $Q_\triangle[a_i]$: 
\begin{enumerate}
\item[(i)] Compute $\sigma_{A=a_i}(R)\Join \sigma_{A=a_i}(T)$ and
  filter the results by probing against $S$ or

\item[(ii)] 
Consider each tuple in $(b,c)\in S$ and check if $(a_i,b)\in R$ and
$(a_i,c)\in T$.
\end{enumerate}

We pick option (i) when $a_i$ is light (low skew) and pick option (ii)
when $a_i$ is heavy (high skew).

\begin{example}
Let us work through the motivating example from
Figure~\ref{fig:triangle-bad}. When we compute $Q_\triangle[a_0]$,
we realize that $a_0$ is heavy and hence, we use option (ii)
above. Since here we just scan tuples in $S$, computing $Q_\triangle[a_0]$ takes
$O(m)$ time. On the other hand, when we want to compute $Q_\triangle[a_i]$ for
$i\ge 1$, we realize that these $a_i$'s are light and so we take
option (i). In these cases $|\sigma_{A=a_i}R|=|\sigma_{A=a_i}T|=1$ 
and hence the algorithm runs in time $O(1)$. As there are $m$ such
light $a_i$'s, the algorithm overall takes $O(m)$ each on the heavy
and light vertices and thus $O(m)=O(N)$ overall which is best possible
since the output size is $\Theta(N)$.
\end{example}

\paragraph*{Algorithm and Analysis} 
Algorithm~\ref{alg:triangle-two} fully specifies how to 
compute $Q_{\triangle}$ using the above idea of two choices. 
Given that the relations $R, S$, and $T$ are already indexed appropriately,
computing $L$ in line~\ref{line:L-triangle} can easily be done in 
time $O(\min\{|R|, |S|, |T|\})$.
Then, for each $a\in L$, the body of the for loop from line~\ref{line:begin-for}
to line~\ref{line:end-for} clearly takes time in the order of
\[ \min\left(|\sigma_{A=a}R|\cdot |\sigma_{A=a}T|,|S|\right),\]
{\em thanks to the power of two choices}!
Thus, the overall time spent by the algorithm is up to constant factors
\begin{equation}
\label{eq:runtime-two}
\sum_{a\in L} \min\left(|\sigma_{A=a}R|\cdot |\sigma_{A=a}T|,|S|\right).
\end{equation}

We bound the sum above by using two inequalities. 
The first is the simple observation that for any $x,y\ge 0$
\begin{equation}
\label{eq:min-ub}
\min(x,y)\le \sqrt{xy}.
\end{equation}
The second is the famous Cauchy-Schwarz inequality\footnote{The inner product of
two vectors is at most the product of their length}:
\begin{equation}
\label{eq:cs}
\sum_{a\in L} x_a\cdot y_a \le \sqrt{\sum_{a\in L} x_a^2}\cdot \sqrt{\sum_{a\in L} y_a^2},
\end{equation}
where $(x_a)_{a\in L}$ and $(y_a)_{a\in L}$ are vectors of real values. 
Applying \eqref{eq:min-ub} to \eqref{eq:runtime-two}, we obtain
\begin{align}
\label{eq:two-sum-orig}
&\sum_{a\in L} \sqrt{|\sigma_{A=a}R|\cdot |\sigma_{A=a}T| \cdot |S|}\\
\label{eq:two-sum}
=&\sqrt{|S|}\cdot \sum_{a\in L} \sqrt{|\sigma_{A=a}R|}\cdot \sqrt{|\sigma_{A=a}T|}\\
\le &\sqrt{|S|}\cdot \sqrt{\sum_{a\in L} |\sigma_{A=a}R|} \cdot \sqrt{\sum_{a\in
L} |\sigma_{A=a}T|}\notag\\
\le &\sqrt{|S|}\cdot \sqrt{\sum_{a\in \pi_A(R)} |\sigma_{A=a}R|} \cdot
\sqrt{\sum_{a\in \pi_A(T)} |\sigma_{A=a}T|}\notag\\
=&\sqrt{|S|}\cdot \sqrt{|R|}\cdot \sqrt{|T|}.\notag
\end{align}
If $|R|=|S|=|T|=N$, then the above is $O(N^{3/2})$ as claimed in the
introduction.
We will generalize the above algorithm beyond triangles to general
join queries in 
Section~\ref{sec:agm-prf-algo}.
Before that, we present a second algorithm that has exactly the same worst-case
run-time and a similar analysis to illustrate the recursive structure of 
the generic worst-case join algorithm described in 
Section~\ref{sec:agm-prf-algo}.

\begin{algorithm}[t]
\caption{Computing $Q_{\triangle}$ with power of two choices.}
\label{alg:triangle-two}
\begin{algorithmic}[1]
\Require{$R(A,B),S(B,C),T(A,C)$ in sorted order}
\Statex
\State $Q_\triangle\gets \emptyset$
\State $L\gets \pi_A(R)\cap \pi_A(T)$ \label{line:L-triangle}
\For{each $a\in L$}
	\If{$|\sigma_{A=a}R|\cdot |\sigma_{A=a}T|\ge |S|$}\label{line:begin-for}
		\For{each $(b,c)\in S$}
			\If{$(a,b)\in R$ and $(a,c)\in T$}
				\State Add $(a,b,c)$ to $Q_\triangle$
			\EndIf
		\EndFor
	\Else
		\For{each $b\in \pi_{B}(\sigma_{A=a}R)\wedge c\in\pi_C(\sigma_{A=a}T)$}
			\If{$(b,c)\in S$}
				\State Add $(a,b,c)$ to $Q_\triangle$
			\EndIf
		\EndFor
	\EndIf\label{line:end-for}
\EndFor
\State \Return{$Q$} 
\end{algorithmic}
\end{algorithm}

\subsection{Algorithm 2: Delaying the Computation} 
Now we present a second way to compute $Q_\triangle[a_i]$ that differentiates
between heavy and light values $a_i\in A$ in a different way. 
We don't try to estimate the heaviness of $a_i$ right off the bat.
Algorithm~\ref{ALG:TRIANGLE-LEAP} ``looks deeper'' into what pair $(b,c)$
can go along with $a_i$ in the output by computing $c$ for each candidate $b$.

Algorithm~\ref{ALG:TRIANGLE-LEAP} works as follows.
By computing the intersection 
$\pi_B(\sigma_{A=a_i}R) \cap \pi_BS$, we only look at the candidates
$b$ that can possibly participate with $a_i$ in the output $(a_i,b,c)$. 
Then, the candidate set for $c$ is
$\pi_{C}(\sigma_{B=b}S) \cap \pi_{C}(\sigma_{A=a_i}T).$
When $a_i$ is really skewed toward the heavy side, the candidates $b$ 
and then $c$ help gradually reduce the skew toward building up the final 
solution $Q_\triangle$.

\iffalse
The power of two choices is again at work in a slightly different way.

\begin{enumerate}
\item If $|\pi_{C}\sigma_{B=b}S| < |\pi_{C}\sigma_{A=a_i}T|$, 
then we compute the intersection by going through all 
$c\in \pi_C\sigma_{B=b}S$ and checking if 
$c\in \pi_C\sigma_{A=a_i}T$.
This process takes time linear in size of the smaller of the two sets. 

\item Otherwise we go through all $c\in \pi_C\sigma_{A=a_i}T$ and check if 
$c\in \pi_C\sigma_{B=b}S$.
\end{enumerate}
\fi

\begin{example}
Let us now see how delaying computation works on the bad
example. 
As we have observed in using the power of two choices, computing the
intersection of two sorted sets takes time at most the {\em minimum} of the
two sizes.

For $a_0$, we consider all $b\in\{b_0,b_1,\dots,b_m\}$. 
When $b=b_0$, we have
\[ \pi_C(\sigma_{B=b_0}S)=\pi_C(\sigma_{A=a_0}T)=\{c_0,\dots,c_m\}, \]
so we output the $m+1$ triangles in total time $O(m)$. 
For the pairs $(a_0,b_i)$ when $i\ge 1$, we have
$|\sigma_{B=b_i}S|=1$ and hence we
spend $O(1)$ time on each such pair, for a total of $O(m)$ overall.

Now consider $a_i$ for $i\ge 1$. In this case,
$b=b_0$ is the only candidate. Further, for $(a_i,b_0)$, we have
$|\sigma_{A=a_i}T|=1$, so we can handle each such $a_i$ in
$O(1)$ time leading to an overall run time of $O(m)$. Thus
on this bad example Algorithm~\ref{ALG:TRIANGLE-LEAP}
runs in $O(N)$ time.
\end{example}

Appendix~\ref{sec:triangle-leap} has the full analysis of
Algorithm~\ref{ALG:TRIANGLE-LEAP}: its worst-case runtime is exactly the same 
as that of Algorithm~\ref{alg:triangle-two}.
What is remarkable is that both of these algorithms follow exactly the same
recursive structure and they are special cases of a generic 
worst-case optimal join algorithm.

\begin{algorithm}[t]
\caption{Computing $Q_{\triangle}$ by delaying computation.}
\label{ALG:TRIANGLE-LEAP}
\begin{algorithmic}[1]
\Require{$R(A,B),S(B,C),T(A,C)$ in sorted order}
\Statex
\State $Q\gets\emptyset$
\State $L_A\gets \pi_A R \cap \pi_A T$
\For{each $a\in L_A$}
	\State $L_B^a\gets \pi_B \sigma_{A=a}R \cap \pi_B S$
	\For{each $b\in L_B^a$}
		\State $L_C^{a,b}\gets \pi_C\sigma_{B=b}S
                \cap \pi_C \sigma_{A=a}T$
		\For{each $c\in L_C^{a,b}$}
			\State Add $(a,b,c)$ to $Q$
		\EndFor
	\EndFor
\EndFor
\State \Return{$Q$}
\end{algorithmic}
%\label{alg:two}
\end{algorithm}

% ----------------------------------------------------------------------------
\section{A User's Guide to the \agm bound}
\label{sec:prelim}
% ----------------------------------------------------------------------------

We now describe one way to generalize the bound of the output size of
a join (mirroring the $O(N^{3/2})$ bound we saw for the triangle
query) and illustrate its use with a few examples.

\subsection{\agm Bound}
\begin{figure*}[t]
  \centering
 \scalebox{.4}{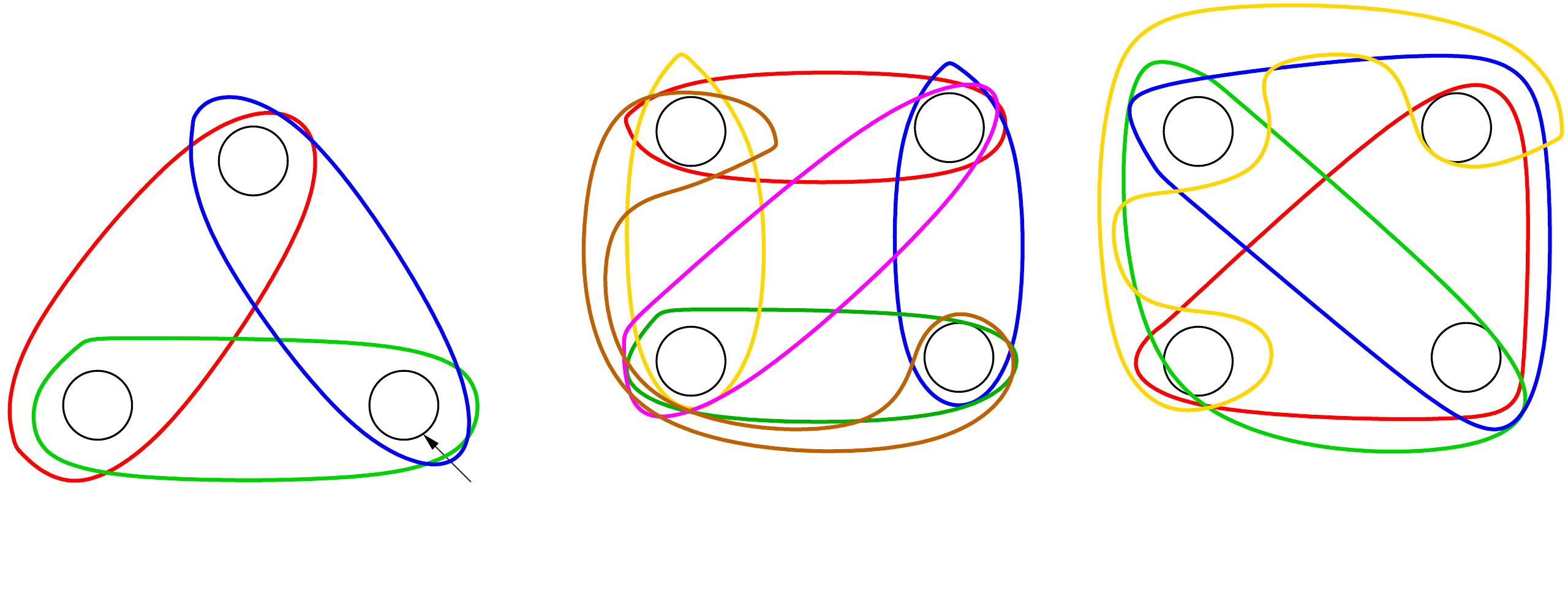}
%\caption{Illustration for bad example for $m=4$.}
\caption{A handful of queries and their covers.}
\label{fig:queries}
\end{figure*}

To state the \agm
bound, we need some notation.
The natural join problem can be defined as follows.  We are given a
collection of $m$ relations.  Each relation is over a collection of
attributes.  We use $\calV$ to denote the set of attributes; let $n =
|\calV|$.  The join query $Q$ is modeled as a hypergraph $\calH =
(\calV, \calE)$, where for each hyperedge $F \in \calE$ there is a
relation $R_F$ on attribute set $F$. Figure~\ref{fig:queries} shows
several example join queries, their associated hypergraphs, and
illustrates the bounds below.

Atserias-Grohe-Marx \cite{AGM08} and Grohe-Marx \cite{GM06}
proved the following remarkable inequality, which shall be referred to as
the {\em \agm's inequality} henceforth. 
Let $\mv x = (x_F)_{F\in\calE}$ be {\em any} point in the following polyhedron: 
\[ \left\{ \mv x \suchthat \sum_{F: v \in F} x_F \geq 1, \forall v \in \calV,
\mv x \geq \mv 0 \right\}. \]
Such a point $\mv x$ is called a {\em fractional edge cover} of the 
hypergraph $\calH$. 
Then, \agm's inequality states that the join size can be bounded by
\begin{equation}
\label{eqn:AGM}
 |Q| = |\Join_{F\in\calE}R_F| \leq \prod_{F\in\calE} |R_F|^{x_F}. 
\end{equation}

\subsection{Example Bounds} 
\label{sec:eg-bound}
%\cmr{We illustrate the bounds above (How they can be loose, how they
%  change with different relation sizes). Also, show how each of the
%  covers in the intro are valid solutions (with slack) for triangles.}
We now illustrate the \agm bound on some specific join queries. We begin with the triangle query $Q_{\triangle}$. In this case the corresponding hypergraph $\cal H$ is as in the left part of Figure~\ref{fig:queries}. We consider two covers (which are also marked in Figure~\ref{fig:queries}). The first one is $x_R=x_T=x_S=\frac{1}{2}$. This is a valid cover since the required inequalities are satisfied for every vertex. For example, for vertex $C$, the two edges incident on it are $S$ and $T$ and we have $x_S+x_T=1\ge 1$ as required. In this case the bound \eqref{eqn:AGM} states that 
\begin{equation}
\label{eq:triang-all-equal}
|Q_{\triangle}|\le \sqrt{|R|\cdot |S|\cdot |T|}.
\end{equation}
Another valid cover is $x_R=x_T=1$ and $x_S=0$ (this cover is also marked in
Figure~\ref{fig:queries}). This is a valid cover, e.g. since for $C$ we have $x_S+x_T=1\ge 1$ and for vertex $A$, we have $x_R+x_T=2\ge 1$ as required. For this cover, bound~\eqref{eqn:AGM} gives
\begin{equation}
\label{eq:triang-two-small}
|Q_{\triangle}| \le |R|\cdot |T|.
\end{equation}
These two bounds can be better in different scenarios. E.g. when $|R|=|S|=|T|=N$, then~\eqref{eq:triang-all-equal} gives an upper bound of $N^{3/2}$ (which is the tight answer) while~\eqref{eq:triang-two-small} gives a bound of $N^2$, which is worse. However, if $|R|=|T|=1$ and $|S|=N$, then~\eqref{eq:triang-all-equal} gives a bound of $\sqrt{N}$, which has a lot of slack; while~\eqref{eq:triang-two-small} gives a bound of $1$, which is tight.

For another class of examples, consider the ``clique" query. In this case there
are $n\ge 3$ attributes and $m=\binom{n}{2}$ relations: one $R_{i,j}$ for every
$i<j\in [n]$: we will call this query $K_n$. Note that $K_3$ is $Q_{\triangle}$.
The middle part of Figure~\ref{fig:queries} considers the $K_4$ query. We
highlight one cover: $x_{R_{i,j}}=\frac{1}{n-1}$ for every $i<j\in [n]$. This is
a valid cover since every attribute is contained in $n-1$ relations. Further, in this case \eqref{eqn:AGM} gives a bound of
$\sqrt[n-1]{\prod_{i<j} |R_{i,j}|},$
which simplifies to $N^{n/2}$ for the case when every relation has size $N$.

Finally, we consider the Loomis-Whitney $LW_n$ queries. In this case there are $n$ attributes and there are $m=n$ relations. In particular, for every 
$i\in [n]$ there is a relation $R_{-i} = R_{[n]\setminus \{i\}}$. Note that 
$LW_3$ is $Q_{\triangle}$. See the right of Figure~\ref{fig:queries} for the $LW_4$ query. We highlight one cover: $x_
{R_{i,j}}=\frac{1}{n-1}$ for every $i<j\in [n]$. This is a valid cover since every attribute is contained in $n-
1$ relations. Further, in this case \eqref{eqn:AGM} gives a bound of
$\sqrt[n-1]{\prod_{i} |R_{-i}|},$
which simplifies to $N^{1+\frac{1}{n-1}}$ for the case when every relation has size $N$. Note that this bound approaches $N$ as $n$ becomes larger.

\subsection{The Tightest \agm Bound}
As we just saw, the optimal edge cover for the \agm bound depends on the relation sizes.
To minimize the right hand side of \eqref{eqn:AGM}, we can solve the
following linear program:
\begin{eqnarray*}
\min && \sum_{F\in \calE} (\log_2 |R_F|) \cdot x_F\\
\text{s.t.} && \sum_{F: v \in F} x_F \geq 1, v \in \calV\\
&&\mv x \geq \mv 0
\end{eqnarray*}
Implicitly, the objective function above depends on the database instance
$\calD$ on which the query is applied. Let $\rho^*(Q, \calD)$ denote the 
optimal objective value to the above linear program. 
We refer to $\rho^*(Q,\calD)$ as the {\em fractional edge cover number} of the 
query $Q$ with respect to the database instance $\calD$,
following Grohe \cite{grohebounds}.
The \agm's inequality can be summarized simply by
$|Q| \leq 2^{\rho^*(Q, \calD)}$.
%\begin{equation}
%\label{eqn:AGMrho}
% |Q| \leq 2^{\rho^*(Q, \calD)}.
%\end{equation}

\subsection{Applying \agm bound on conjunctive queries with
simple functional dependencies}
\label{subsec:conjunctive}
% ----------------------------------------------------------------------------

Thus far we have been describing bounds and algorithms for natural
join queries. A super-class of natural join queries called {\em
  conjunctive queries}. A conjunctive query is a query of the form
\[ C = R_0(\bar X_0) \gets R_1(\bar X_1) \wedge \cdots \wedge R_m(\bar X_m) \]
where 
$\{R_1,\dots,R_m\}$ is a multi-set of relation symbols, i.e. some 
relation might occur more than once in the query, and
$\bar X_0, \dots, \bar X_m$ are tuples of variables, and each variable
occurring in the query's head $R(\bar X_0)$ must also occur in the body. It is
important to note that the same variable might occur more than once in the 
same tuple $\bar X_i$.

We will use $\vars(C)$ to denote the set of all variables occurring in $C$.
Note that $\bar X_0 \subseteq \vars(C)$ and it is entirely possible for
$\bar X_0$ to be {\em empty} (Boolean conjunctive query).
For example, the following are conjunctive queries:
\begin{eqnarray*}
R_0(WXYZ) &\gets& S(WXY) \wedge S(WWW) \wedge T(YZ)\\
R_0(Z)    &\gets& S(WXY) \wedge S(WWW) \wedge T(YZ).
\end{eqnarray*}
The former query is a {\em full conjunctive query} because the head atom 
contains all the query's variable.

Following Gottlob, Lee, Valiant, and Valiant (\glvv hence forth)
\cite{GLVV09, GLVV09-conf}, we also know that the \agm bound can be
extended to general conjunctive queries even with simple functional
dependencies.\footnote{\glvv also have fascinating bounds for the
  general functional dependency and composite keys cases, and
  characterization of treewidth-preserving queries; both of those
  topics are beyond the scope of this survey, in particular because
  they require different machinery from what we have developed thus
  far.} In this survey, our presentation closely follows Grohe's
presentation of \glvv~\cite{grohebounds}.

To illustrate what can go ``wrong'' when we are moving from natural
join queries to conjunctive queries, let us first consider a few example
conjunctive queries, introducing one issue at a time.
In all examples below, relations are assumed to have the same size $N$.

\begin{example}[Projection]
Consider 
\[ C_1 = R_0(W) \gets R(WX) \wedge S(WY) \wedge T(WZ).\]
 In the (natural)
 join query, $R(WX) \wedge S(WY) \wedge T(WZ)$ \agm bound gives $N^3$; but 
 because $R_0(W) \subseteq \pi_W(R) \Join \pi_W(S) \Join \pi_W(T)$, \agm bound
 can be adapted to the instance restricted only to the output variables
 yielding an upper bound of $N$ on the output size.
\label{ex:projection}
\end{example}

\begin{example}[Repeated variables] 
Consider the query
\[ C_2 = R_0(WY) \gets R(WW) \wedge S(WY) \wedge T(YY).\] 
This is a full conjunctive query as all variables appear in the head 
atom $R_0$. In this case, we can replace $R(WW)$ and $T(YY)$ by keeping 
only tuples $(t_1,t_2) \in R$ for which $t_1=t_2$ and
tuples $(t_1,t_2)\in T$ for which $t_1=t_2$; essentially, we turn the query
into a natural join query of the form $R'(W) \wedge S(WY) \wedge T'(Y)$.
For this query, $x_{R'}=x_{T'}=0$ and $x_S=1$ is a fractional cover
and thus by \agm bound $N$ is an upperbound on the output size.
\label{ex:repeated}
\end{example}

\begin{example}[Introducing the chase]
\label{ex:chase}
Consider the query
\[ C_3 = R_0(WXY) \gets R(WX) \wedge R(WW) \wedge S(XY). \]
 Without additional information, the best bound we can get for this query 
 is $O(N^2)$: we can easily turn it into a natural join query of the form
 $R(WX) \wedge R'(W) \wedge S(XY)$, where $R'$ is obtained from $R$ by keeping
 all tuples $(t_1,t_2) \in R$ for which $t_1=t_2$. 
 Then, $(x_R, x_{R'}, x_S)$ is a fractional edge cover for this query if and only if
 $x_R+x_{R'} \geq 1$ (to cover $W$), 
 $x_R+x_{S} \geq 1$ (to cover $X$), 
 $x_S \geq 1$ (to cover $Y$);
 So, $x_S=x_{R'}=1$ and $x_R=0$ is a fractional cover, yielding the $O(N^2)$
 bound.
 Furthermore, it is easy to construct input instances for 
 which the output size is $\Omega(N^2)$:
 \begin{eqnarray*}
 R &=& \{ (i, i) \suchthat i \in [N/2] \} \bigcup \{ (i, 0) \suchthat i \in
 [N/2] \}\\
 S &=& \{ (0, j) \suchthat j \in [N] \}.
 \end{eqnarray*}
 Every tuple $(i, 0, j)$ for $i\in [N/2], j\in [N]$ is in the output.

 Next, suppose we have an additional piece of information that the first 
 attribute in relation $R$ is 
 its {\em key}, i.e. if $(t_1,t_2)$ and $(t_1,t'_2)$ are in $R$, then $t_2 =
 t'_2$. Then we can significantly reduce the output size bound because we 
 can infer the following about the output tuples: $(w,x,y)$ is an output
 tuple iff $(w,x)$ and $(w,w)$ are in $R$, and $(x,y)$ are in $S$.
 The functional dependency tells us that $x=w$. Hence, the query is equivalent
 to 
 \[ C'_3 = R_0(WY) \gets R(WW) \wedge S(WY).\]
 The \agm bound for this (natural) join query is $N$. The transformation from
 $C_3$ to $C'_3$ we just described is, of course, the famous {\em chase}
 operation \cite{Maier:1979:TID:320107.320115, DBLP:journals/tods/AhoBU79,
 DBLP:journals/jacm/BeeriV84}, which is much more powerful than what conveyed
 by this example.
\end{example}

\begin{example}[Taking advantage of FDs]
Consider the following query
\[ C_4 = R_0(XY_1,\dots,Y_k, Z) \gets \bigwedge_{i=1}^k R_i(XY_i) \wedge 
                        \bigwedge_{i=1}^k S_i(Y_iZ). 
\]
First, without any functional dependency, \agm bound gives $N^k$ for this 
query, because the fractional cover constraints are
\begin{eqnarray*}
\sum_{i=1}^k x_{R_i} & \geq & 1 \text { (cover $X$)}\\
x_{R_i} + x_{S_i} & \geq & 1 \text { (cover $Y_i$) } i \in [k]\\
\sum_{i=1}^k x_{S_i} & \geq & 1 \text { (cover $Z$)}.
\end{eqnarray*}
The \agm bound is $N^{\sum_i(x_{R_i}+x_{S_i})} \geq N^k$.

Second, suppose we know $k+1$ functional dependencies: each of the first 
attributes of relations $R_1,\dots,R_k$ is a key for the corresponding
relation, and the first attribute of $S_1$ is its key. Then, we have the
following sets of functional dependencies:
$X \to Y_i$, $i\in [k]$, and $Y_1 \to Z$.
Now, construct a fictitious relation $R'(X, Y_1,\dots,Y_k,Z)$
as follows: $(x,y_1,\dots,y_k,z) \in R'$ iff $(x,y_i) \in R_i$ for all
$i \in [k]$ and $(y_1,z)\in S_1$. Then, obviously $|R'| \leq N$. More
importantly, the output does not change if we add $R'$ to the body
query $C_4$ to obtain a new conjunctive query $C'_4$.
However, this time we can set $x_{R'}=1$ and all other variables in the
fractional cover to be $0$ and obtain an upper bound of $N$.
\label{ex:FD}
\end{example}

We present a more formal treatment of the steps needed to convert a conjunctive query with simple functional dependencies to a join query in Appendix~\ref{app:glvv}.

% ----------------------------------------------------------------------------
\section{Worst-case-optimal algorithms}
\label{sec:agm-prf-algo}
% ----------------------------------------------------------------------------

We first show how to analyze the upper bound that proves \agm and from
which we develop a generalized join algorithm that captures both algorithms
from Ngo-Porat-R\'e-Rudra~\cite{NPRR} (henceforth \nprr) and 
\leapfrog~\cite{leapfrog}. Then,
we describe the limitation of any join-project plan.

Henceforth, we need the following notation.
Let $\calH=(\calV,\calE)$ be any hypergraph and 
$I\subseteq\calV$ be an arbitrary
subset of vertices of $\calH$. Then, we define
\[ \calE_I := \left\{ F\in\calE \suchthat F\cap I \neq \emptyset\right\}. \]

\begin{example}
For the query $Q_{\triangle}$ from Section~\ref{sec:triangle}, we have 
$\calH_{\triangle}=(\calV_\triangle, \calE_\triangle)$, where
\begin{eqnarray*}
\calV_{\triangle} &= &\{A,B,C\},\\
\calE_{\triangle} &=&\bigl\{\{ A,B\}, \{B,C\}, \{A,C\}\bigr\}. 
\end{eqnarray*}
Let $I_1=\{A\}$ and $I_2=\{A,B\}$, then 
$\calE_{I_1}=\{\{A,B\}, \{A,C\}\}$,
and
$\calE_{I_2}=\calE_\triangle$.
\end{example}

\subsection{A proof of the \agm bound}

We prove the \agm inequality in two steps: a query decomposition lemma,
and then a succinct inductive proof, which we then use to develop a
generic worst-case optimal join algorithm.

\subsubsection{The query decomposition lemma}
\label{subsec:keylemma}

Ngo-Porat-R\'e-Rudra \cite{NPRR} 
gave an inductive proof of \agm bound 
\eqref{eqn:AGM} using H\"older inequality. (\agm proved the bound using an entropy based argument: see Appendix~\ref{app:entropy} for more details.)
The proof has an inductive structure leading naturally
to recursive join algorithms.
\nprr's strategy is a generalization of 
the strategy in \cite{MR1338683} to 
prove the Bollob\'as-Thomason inequality,
shown in \cite{NPRR} to be {\em equivalent} to \agm's bound.

Implicit in \nprr is the following key lemma, which will be crucial in 
proving bounds on general join queries (as well as proving upper bounds on 
the runtime of the new join algorithms).

\blmm[Query decomposition lemma]\label{lmm:Qdecomp}
Let $Q = \ \Join_{F\in\calE} R_F$ be a natural join query represented by a 
hypergraph $\calH=(\calV,\calE)$, and 
$\mv x$ be any fractional edge cover for $\calH$.
Let $\calV = I \uplus J$ be an arbitrary partition of $\calV$ such 
that $1\leq|I|<|\calV|$; and,
\[ L = \Join_{F \in \calE_I} \pi_{I}(R_F). \]
Then,
\begin{equation}
 \sum_{\mv t_I \in L} \prod_{F\in\calE_{J}} |R_F \lJoin \mv t_I|^{x_F}
 \leq \prod_{F\in\calE}|R_F|^{x_F}.
\label{eqn:keyinequality}
\end{equation}
\elmm

Before we prove the lemma above, we outline how we have already used the lemma
above specialized to $Q_{\triangle}$ in Section~\ref{sec:triangle} to bound the
runtime of Algorithm~\ref{alg:triangle-two}. 
We use the lemma with $\mv x=(1/2,1/2,1/2)$, which is a valid fractional edge 
cover for $\calH_{\triangle}$. 
%Further, note that the RHS 
%of~\eqref{eqn:keyinequality} if $N^{3/2}$ for the case of $|R|=|T|=|S|=N$ 
%(which is what we considered in Section~\ref{sec:triangle}).

For Algorithm~\ref{alg:triangle-two} we use Lemma~\ref{lmm:Qdecomp} with 
$I=I_1$. Note that $L$ in Lemma~\ref{lmm:Qdecomp} is the same as 
\[\pi_A(R)\Join \pi_A(T)=\pi_A(R)\cap \pi_A(T),\]
i.e. this $L$ is exactly the same as the $L$ in
Algorithm~\ref{alg:triangle-two}. We now consider the left hand side (LHS)
in~\eqref{eqn:keyinequality}. Note that we have 
$\calE_J=\bigl\{\{A,B\}, \{B,C\}, \{A,C\}\bigr\}$. Thus, the LHS is the 
same as
\begin{align*}
&\sum_{a\in L} \sqrt{|R\lJoin (a)|}\cdot\sqrt{|T\lJoin (a)|} \cdot\sqrt{|S\lJoin (a)|}\\
%=&\sum_{a\in L} \sqrt{|\sigma_{A=a}(R)|}\cdot \sqrt{|\sigma_{A=a}(T)|}\cdot \sqrt{|S|}\\
=&\sum_{a\in L} \sqrt{|\sigma_{A=a}R|}\cdot \sqrt{|\sigma_{A=a}T|}\cdot \sqrt{|S|}.
\end{align*}
Note that the last expression is exactly the same as~\eqref{eq:two-sum-orig},
which is at most $\sqrt{|R|\cdot |S|\cdot |T|}$ by
Lemma~\ref{lmm:Qdecomp}.
This was what shown in Section~\ref{sec:triangle}.

\bp[Proof of Lemma~\ref{lmm:Qdecomp}]
The plan is to ``unroll'' the sum of products on the left hand side using
H\"older inequality as follows.
Let $j \in I$ be an arbitrary attribute. 
Define 
\begin{eqnarray*}
I' &=& I-\{j\}\\
J' &=& J\cup\{j\}\\
L' &=& \Join_{F\in \calE_{I'}} \pi_{I'}(R_F).
\end{eqnarray*}
We will show that
\begin{equation}
 \sum_{\mv t_I \in L} \prod_{F\in\calE_{J}} |R_F \lJoin \mv t_I|^{x_F}
 \leq
 \sum_{\mv t_{I'} \in L'} \prod_{F\in\calE_{J'}} |R_F \lJoin \mv t_{I'}|^{x_F}.
\label{eqn:L'}
\end{equation}
Then, by repeated applications of \eqref{eqn:L'} we will bring $I'$ down
to empty and the right hand side is that of \eqref{eqn:keyinequality}.

To prove \eqref{eqn:L'} we write $\mv t_I = (\mv t_{I'}, t_j)$ for some
$\mv t_{I'} \in L'$ and decompose a sum over $L$ to a double sum over $L'$
and $t_j$, where the second sum is only over $t_j$ for which 
$(\mv t_{I'},t_j) \in L$.
{\allowdisplaybreaks
\begin{eqnarray*}
\sum_{\mv t_I \in L} \prod_{F\in\calE_J} |R_F \lJoin \mv t_I|^{x_F}
&=& \sum_{\mv t_{I'} \in L'} \sum_{t_j}
    \prod_{F\in\calE_J} |R_F \lJoin (\mv t_{I'}, t_j)|^{x_F}\\
&=& \sum_{\mv t_{I'} \in L'} \sum_{t_j}
    \left(\prod_{F\in\calE_J} |R_F \lJoin (\mv t_{I'}, t_j)|^{x_F}\right) \cdot
    \left(\prod_{F\in\calE_{J'}-\calE_J} 1^{x_F}\right) \\
&=& \sum_{\mv t_{I'} \in L'} \sum_{t_j}
    \prod_{F\in\calE_{J'}} |R_F \lJoin (\mv t_{I'}, t_j)|^{x_F}\\
&=& \sum_{\mv t_{I'} \in L'} 
    \prod_{F\in\calE_{J'}-\calE_{\{j\}}} |R_F \lJoin \mv t_{I'}|^{x_F}
    \sum_{t_j}
    \prod_{F\in\calE_{\{j\}}} |R_F \lJoin (\mv t_{I'}, t_j)|^{x_F}\\
&\leq& \sum_{\mv t_{I'} \in L'} 
    \prod_{F\in\calE_{J'}-\calE_{\{j\}}} |R_F \lJoin \mv t_{I'}|^{x_F}
    \prod_{F\in\calE_{\{j\}}} 
    \left(\sum_{t_j}
    |R_F \lJoin (\mv t_{I'}, t_j)|\right)^{x_F}\\
&\leq& \sum_{\mv t_{I'} \in L'} 
    \prod_{F\in\calE_{J'}-\calE_{\{j\}}} |R_F \lJoin \mv t_{I'}|^{x_F}
    \prod_{F\in\calE_{\{j\}}} 
    |R_F \lJoin \mv t_{I'}|^{x_F}\\
&=& \sum_{\mv t_{I'} \in L'} 
    \prod_{F\in\calE_{J'}} |R_F \lJoin \mv t_{I'}|^{x_F}.
\end{eqnarray*}
}
In the above, the third equality follows from fact that for any $F\in\calE_{J'}-\calE_J$, we have $F\subseteq I'\cup \{j\}$. The first inequality is an application of H\"older inequality,
which holds because $\sum_{F\in\calE_{\{j\}}} x_F \geq 1$.
The second inequality is true because the sum is 
only over $t_j$ for which $(\mv t_{I'}, t_j) \in L$.
\ep

It is quite remarkable that from the query decomposition lemma, we can prove 
\agm inequality \eqref{eqn:AGM}, and describe and analyze two join algorithms 
succinctly.

\subsubsection{An inductive proof of \agm inequality}
\label{subsec:new-agm}

{\bf Base case.}
In the base case $|\calV|=1$, we are computing the join of $|\calE|$ unary
relations.
Let $\mv x = (x_F)_{F\in\calE}$ be a fractional edge cover for this instance.
%Then $\mv x \geq 0$ and 
%\[ \sum_{F\in\calE} x_F \geq 1. \]
%\agm bound states that
%\[ | \Join_{F\in\calE} R_F| = |\bigcap_{F\in\calE} R_F|
%   \leq \prod_{F\in\calE} |R_F|^{x_F}.
%\]
%This bound can be shown by noting that
Then,
\begin{eqnarray*}
 |\Join_{F\in\calE} R_F|
 &\leq& \min_{F\in\calE} |R_F|\\
 &\leq& \left(\min_{F\in\calE} |R_F|\right)^{\sum_{F\in\calE} x_F}\\
 &=& \prod_{F\in\calE} \left(\min_{F\in\calE} |R_F|\right)^{x_F}\\
 &\leq& \prod_{F\in\calE} |R_F|^{x_F}.
\end{eqnarray*}

{\bf Inductive step.} Now, assume $n = |\calV| \geq 2$. 
Let $\calV = I \uplus J$ be any partition of $\calV$ such that
$1\leq|I|<|\calV|$. Define $L = \ \Join_{F\in\calE_I} \pi_I(R_F)$ as in
Lemma~\ref{lmm:Qdecomp}. 
For each tuple $\mv t_I \in L$ we define a new join query
\[ Q[\mv t_I] := \Join_{F\in \calE_J} \pi_J(R_F \lJoin \mv t_I).  \]
Then, obviously we can write the original query $Q$ as
\begin{equation}
 Q = \bigcup_{\mv t_I \in L} \left(\{\mv t_I\} \times Q[\mv t_I]\right).
\label{eqn:recQ}
\end{equation}
The vector $(x_F)_{F\in\calE_J}$ is a fractional edge cover for the hypergraph
of $Q[\mv t_I]$. Hence, the induction hypothesis gives us
\begin{equation}
 |Q[\mv t_I]| \leq \prod_{F\in\calE_J} |\pi_J(R_F \lJoin \mv t_I)|^{x_F}
              = \prod_{F\in\calE_J} |R_F \lJoin \mv t_I|^{x_F}. 
\label{eqn:indhypo}
\end{equation}
From \eqref{eqn:recQ}, \eqref{eqn:indhypo}, and \eqref{eqn:keyinequality} we obtain
\agm inequality:
\[ |Q| = \sum_{\mv t_I \in L} |Q[\mv t_I]| \leq \prod_{F\in\calE} |R_F|^{x_F}. 
\]

\subsection{Worst-case optimal join algorithms}
\label{subsec:generic-join}

From the proof of Lemma~\ref{lmm:Qdecomp} and the query decomposition
\eqref{eqn:recQ}, it is straightforward to design a class of recursive 
join algorithms which is optimal in the worst case. (See
Algorithm~\ref{alg:generic-join}.)

\begin{algorithm}[th]
\caption{Generic-Join($\Join_{F\in\calE} R_F$)}
\label{alg:generic-join}
\begin{algorithmic}[1]
\Require{Query $Q$, hypergraph $\calH=(\calV,\calE)$}
\Statex
\State $Q \gets \emptyset$
\If {$|\calV|=1$} 
  \State return $\bigcap_{F\in\calE} R_F$
\EndIf
\State Pick $I$ arbitrarily such that $1\leq|I|<|\calV|$
\label{step:generic:choice-I}
\State $L \gets$ Generic-Join$(\Join_{F\in\calE_I} \pi_I(R_F))$
\label{step:generic:L}
\For{every $\mv t_I \in L$}
\label{step:generic-last-call}
	\State $Q[\mv t_I] \gets $ Generic-Join$(\Join_{F\in \calE_J} \pi_J(R_F
    \lJoin \mv t_I))$
    \State $Q \gets Q \cup \{\mv t_I\} \times Q[\mv t_I]$
\EndFor
\State \Return{$Q$}
\end{algorithmic}
\end{algorithm}

A mild assumption which is not very crucial is to pre-index all the relations
so that the inputs to the sub-queries $Q[\mv t_I]$ can readily be available
when the time comes to compute it. Both \nprr and \leapfrog 
algorithms do this by fixing a global attribute order and build a B-tree-like
index structure for each input relation consistent with this global
attribute order.  \nprr also described an hash-based indexing structure so as
to remove a $\log$-factor from the final run time. We will not delve on this
point here, except to emphasize the fact that we do not include the linear
time pre-processing step in the final runtime formula.

Given the indices, when $|\calV|=1$ computing $\bigcap_{F\in\calE}
R_F$ can easily be done in time
\[ \tilde O(m \min|R_F|) = \tilde O(m \prod_{F\in\calE}|R_F|^{x_F}). \]
Then, given this base-case runtime guarantee, we can show by induction
that the overall runtime of Algorithm~\ref{alg:generic-join} 
is $\tilde O( mn \prod_{F\in\calE}|R_F|^{x_F})$, where $\tilde O$ hides
a potential $\log$-factor of the input size.
This is because, by induction the time it takes to compute $L$ is
$\tilde O( m|I| \prod_{F\in\calE_I} |R_F|^{x_F} )$, and the time
it takes to compute $Q[\mv t_I]$ is
\[ \tilde O\left( m(n-|I|) \prod_{F\in\calE_J} |R_F \lJoin \mv t_I|^{x_F}
   \right)
\]
Hence, from Lemma~\ref{lmm:Qdecomp}, the total run time is $\tilde O$ of
\begin{eqnarray*}
&&m|I| \prod_{F\in\calE_I} |R_F|^{x_F} +
m(n-|I|)\sum_{\mv t_I\in L} \prod_{F\in\calE_J} |R_F \lJoin \mv t_I|^{x_F}\\
&\leq& m|I| \prod_{F\in\calE_I} |R_F|^{x_F} +
m(n-|I|) \prod_{F\in\calE} |R_F|^{x_F}\\
&\leq& mn \prod_{F\in\calE} |R_F|^{x_F}.
\end{eqnarray*}

The \nprr algorithm is an instantiation of Algorithm~\ref{alg:generic-join} 
where it picks $J \in \calE$, $I = \calV-J$, and solves the sub-queries 
$Q[\mv t_I]$ in a different way, making use of the power of two choices idea. 
Since $J\in \calE$, we write
\[ Q[\mv t_I] = \ R_J \Join 
   \left(\Join_{F\in \calE_J-\{J\}} \pi_J(R_F \lJoin \mv t_I)\right).
\]
Now, if $x_J \geq 1$ 
then we solve for $Q[\mv t_I]$ by checking for every tuple in $R_J$ whether
it can be part of $Q[\mv t_I]$. The run time is $\tilde O$ of
\[ (n-|I|)|R_J| \leq (n-|I|) \prod_{F\in\calE_J} |R_F \lJoin \mv t_I|^{x_F}.
\]
When $x_J < 1$, we will make use of an extremely simple observation:
for any real numbers $p, q \geq 0$ and $z \in [0,1]$, 
$\min\{p,q\} \leq p^zq^{1-z}$ (note that \eqref{eq:min-ub} is the special case of $z=1/2$). In particular,
define
\begin{eqnarray*}
p &=& |R_J|\\
q &=& \prod_{F\in\calE_J-\{J\}} |\pi_J(R_F \lJoin \mv t_I)|^{\frac{x_F}{1-x_J}}
\end{eqnarray*}
Then,
\begin{eqnarray*}
\min \left\{ p, q \right\}
&\leq& |R_J|^{x_J} \prod_{F\in\calE_J-\{J\}} |\pi_J(R_F \lJoin \mv t_I)|^{x_F}\\
&=& \prod_{F\in\calE_J} |R_F \lJoin \mv t_I|^{x_F}.
\end{eqnarray*}

From there, when $x_J<1$ and $p\leq q$, 
we go through each tuple in $R_J$ and check as in the case $x_J\geq 1$.
And when $p>q$, we solve the subquery 
$\Join_{F\in\calE_J-\{J\}} \pi_J(R_F \lJoin \mv t_I)$
first using $\left( \frac{x_F}{1-x_J} \right)_{F\in\calE_J - \{J\}}$ as
its fractional edge cover;
and then check for each tuple in the result whether it is in $R_J$.
In either case, the run time $\tilde O(\min\{p,q\})$ which
along with the observation above completes the proof.

Next we outline how Algorithm~\ref{alg:triangle-two} is
Algorithm~\ref{alg:generic-join} with the above modification for \nprr for 
the triangle query $Q_{\triangle}$. In particular, we will use 
$\mv x=(1/2,1/2,1/2)$ and $I=\{A\}$. Note that this choice of $I$ implies 
that $J=\{B,C\}$, which means in Step~\ref{step:generic:L} 
Algorithm~\ref{alg:generic-join} computes 
\[L=\pi_A(R)\Join \pi_A(T)=\pi_A(R)\cap \pi_A(T),\]
which is exactly the same $L$ as in Algorithm~\ref{alg:triangle-two}. Thus, in the remaining part of Algorithm~\ref{alg:generic-join} one would cycle through all $a\in L$ (as one does in Algorithm~\ref{alg:triangle-two}). In particular, by the discussion above, since $x_{S}=1/2<1$, we will try the best of two 
choices. In particular, we have
\begin{eqnarray*}
\Join_{F\in\calE_J-\{J\}} \pi_J(R_F \lJoin \mv (a))
&=&\pi_B(\sigma_{A=a}R) \times \pi_C(\sigma_{A=a}T),\\
p&=&|I|,\\
q&=& |\sigma_{A=a}R|\cdot |\sigma_{A=a}T|.
\end{eqnarray*}
Hence, the \nprr algorithm described exactly matches 
Algorithm~\ref{alg:triangle-two}.

The \leapfrog algorithm \cite{leapfrog} is an instantiation of 
Algorithm~\ref{alg:generic-join} where $\calV = [n]$ and 
$I = \{1,\dots,n-1\}$ (or equivalently $I=\{1\}$). 
To illustrate, we outline how 
Algorithm~\ref{ALG:TRIANGLE-LEAP} is Algorithm~\ref{alg:generic-join} with 
$I=\{A,B\}$ when specialized to $Q_{\triangle}$. Consider the run of 
Algorithm~\ref{alg:generic-join} on $\calH_{\triangle}$, and the first time 
Step~\ref{step:generic:choice-I} is executed. The call to Generic-Join in 
Step~\ref{step:generic:L} returns $L=\{(a,b)|a\in L_A, b\in L_B^a\}$, 
where $L_A$ and $L_B^a$ are as defined in Algorithm~\ref{ALG:TRIANGLE-LEAP}. 
The rest of Algorithm~\ref{alg:generic-join} is to do the following for 
every $(a,b)\in L$. $Q[(a,b)]$ is computed by the recursive call to 
Algorithm~\ref{alg:generic-join} to obtain $\{(a,b)\times L^{a,b}_C$, where 
\[
L^{a,b}_C = \pi_C\sigma_{B=b}S\Join \pi_C\sigma_{A=a}T
= \pi_C\sigma_{B=b}S\cap \pi_C\sigma_{A=a}T,
\]
exactly as was done in Algorithm~\ref{ALG:TRIANGLE-LEAP}. Finally, we get back to $L$ in Step~\ref{step:generic:L} being as claimed above. Note that in during the recursive call of Algorithm~\ref{alg:generic-join} on $Q_{\bowtie}=R\Join \pi_B(S)\Join \pi_A(T)$. The claim follows by picking $I=\{A\}$ in Step~\ref{step:generic:choice-I} when Algorithm~\ref{alg:generic-join} is run on $Q_{\bowtie}$ (and tracing through rest of Algorithm~\ref{alg:generic-join}).

\subsection{On the limitation of any join-project plan}

\agm proved that there are classes of queries for
which join-only plans are significantly worse than their join-project plan.
In particular, they showed that for every $M, N \in \mathbb N$, there is a query
$Q$ of size at least $M$ and a database $\calD$ of size at least $N$ such that
$2^{\rho^*(Q,\calD)} \leq N^2$ and every join-only plan runs in time at least 
$N^{\frac 1 5 \log_2|Q|}$.

\nprr continued with the story and noted that for the class of $LW_n$ queries from Section~\ref{sec:eg-bound}
every join-project plan runs in time polynomially worse than
the \agm bound.
%\footnote{We thank an anonymous PODS'12 referee for sketching 
%to us the argument showing that our example works for all join-project plans 
%rather than just the \agm algorithm and a join-tree algorithm.}
%We briefly describe this class of queries here.
%
%Call a query $Q$ as a {\em Loomis-Whitney query} (or {\em LW query}
%for short) if its corresponding hypergraph $\calH=(\calV,\calE)$
%has the following structure:
%$\calV=[n]$, and $\calE = \binom{[n]}{n-1}$ for some integer $n \geq 2$.
The proof of the following lemma can be found in 
Appendix~\ref{sec:bad-instance}.

\blmm
Let $n\geq 2$ be an arbitrary integer.
For any LW-query $Q$ with corresponding hypergraph 
$\calH = ([n], \binom{[n]}{n-1})$, and any positive integer $N\geq 2$, there 
exist $n$ relations $R_i$, $i\in[n]$
such that $|R_i| =N,\forall i\in [n]$, the attribute set for $R_i$ is
$[n]-\{i\}$, and that {\em any} join-project plan for $Q$ on these relations
has run-time at least $\Omega(N^2/n^2)$.
\label{LEM:BAD:INSTANCE}
\elmm

Note that both the traditional join-tree-based algorithms and \agm's 
algorithm described in Appendix~\ref{subsec:AGMproof} are join-project 
plans.  Consequently, they run in time asymptotically worse than the best 
\agm bound for this instance, which is
\[ |\Join_{i=1}^n R_i|\le \prod_{i=1}^n |R_i|^{1/(n-1)}=N^{1+1/(n-1)}. \]

On the other hand, both algorithms described in 
Section~\ref{subsec:generic-join} 
take $O(N^{1 + 1/(n-1)})$-time because their run times 
match the \agm bound.
In fact, the \nprr algorithm in Section~\ref{subsec:generic-join} can be shown 
to run in linear data-complexity time $O(n^2N)$ for this query
\cite{NPRR}.

\section{Open Questions}
\label{sec:concl}
% ----------------------------------------------------------------------------

We conclude this survey with two open questions: one for systems
researchers and one for theoreticians:

\begin{enumerate}
\item A natural question to ask is whether the algorithmic ideas that
  were presented in this survey can gain runtime efficiency in
  databases systems. This is an intriguing open question: on one hand
  we have shown asymptotic improvements in join algorithms, but on the
  other there are several decades of engineering refinements and
  research contributions in the traditional dogma.

\item Worst-case results, as noted by several authors, may only give
  us information about pathological instances. Thus, there is a
  natural push toward more refined measures of complexity. For
  example, current complexity measures are too weak to explain why
  indexes are used or give insight into the average case. For example,
  could one design an adaptive join algorithm whose run time is
  somehow dictated by the ``difficulty" of the input instance (instead
  of the input size as in the currently known results)?
\end{enumerate}

\section*{Acknowledgements}                                                     
                                                                                
HN's work is partly supported by NSF grant CCF-1319402 and a gift from          
Logicblox.                                                                      
CR's work on this project is generously supported by NSF CAREER Award under     
No. IIS-1353606, NSF award under No. CCF-1356918, the ONR                       
under awards No.  N000141210041 and No. N000141310129,                          
Sloan Research Fellowship, Oracle, and Google.                                  
AR's work is partly supported by                                                
NSF CAREER Award CCF-CCF-0844796, NSF grant CCF-1319402 and a                   
gift from Logicblox. 

%\bibliographystyle{acm}
%%
%{%\scriptsize
%\bibliography{main,references-hqn}
%}

\def\cprime{$'$} \def\shortbib{0}

\appendix

\section{Relation Algebra Notation}
\label{app:notation}

We assume the existence of a set of attribute names $\mathcal{A} =
A_1,\dots, A_n$ with associated domains $\D_1,\dots,\D_n$ and infinite
set of relational symbols $\mathcal R$. 
A relational schema for the symbol $R \in \mathcal R$ of arity $k$ is a 
tuple $\bar A(R) = (A_{i_1}, \dots, A_{i_k})$ of distinct attributes that 
defines the attributes of the relation. 
A relational database schema is a set of relational symbols
and associated schemas denoted by $R(\bar A(R)), R\in \mathcal R$.
A relational instance for $R(A_{i_1},\dots,A_{i_k})$ is a
subset of $\D_{i_1} \times \dots \times \D_{i_k}$. A relational database 
$\calD$ is a collection of instances, one for each relational symbol in 
schema, denoted by $R^{\calD}$.

A {\em natural join} query (or simply join query) $Q$ is
specified by a finite subset of relational symbols $\atoms(Q) \subseteq
\mathcal R$, denoted by $\Join_{R \in \atoms(Q)} R$. Let $\bar A(Q)$ denote
the set of all attributes that appear in some relation in $Q$, that is
$\bar A(Q) = \{ A \suchthat A \in \bar A(R) \text{ for some } 
R \in \atoms(Q)\}$.  
Given a tuple
$\mv t$ we will write $\mv t_{\bar A}$ to emphasize that its support
is the attribute set $\bar A$. Further, for any $\bar S\subset \bar A$
we let $\mv t_{\bar S}$ denote $\mv t$ restricted to $\bar S$.
Given a database instance $\calD$, the
output of the query $Q$ on the database instance $\calD$ is denoted 
$Q(\calD)$ and is defined as
\[ Q(\calD) \stackrel{\mathrm{def}}{=} \left\{ \mv t \in \D^{\bar A(Q)} 
\suchthat 
\mv t_{\bar A(R)} \in R^{\calD} \text{ for each } R \in \atoms(Q)\right\} \]
where $\D^{\bar A(Q)}$ is a shorthand for 
$\times_{i : A_i \in \bar A(Q)} \D_i$. When the instance is clear from the 
context we will refer to $Q(\calD)$ by just $Q$.

We also use the notion of a {\em semijoin}: Given two relations
$R(\bar A)$ and $S(\bar B)$ their semijoin $R \lJoin S$  is defined by
\[ R \lJoin S \stackrel{\mathrm{def}}{=} 
  \left\{ \mv t \in R : \exists \mv u \in S \text{ s.t. } 
\mv t_{\bar A \cap \bar B} = \mv u_{\bar A \cap \bar B} \right\}.
\]
For any relation $R(\bar A)$, and any subset $\bar S\subseteq \bar A$
of its attributes, let $\pi_{\bar S}(R)$ denote the {\em projection} of 
$R$ onto $\bar S$, i.e.
\[ \pi_{\bar S}(R) = \left\{\mv t_{\bar S} \suchthat
   \exists \mv t_{\bar A\setminus \bar S}, (\mv t_{\bar S}, \mv t_{\bar A\setminus \bar S})
   \in R \right\}.
\] 

For any relation $R(\bar A)$, any subset $\bar S\subseteq \bar A$
of its attributes, and a vector $\mv s\in\prod_{i\in \bar S} \D_i$, let $\sigma_{\bar S=\mv s}(R)$ denote the {\em selection} of 
$R$ with $\bar S$ and $\mv s$, i.e.
\[ \sigma_{\bar S=\mv s}(R)=\left\{ \mv t\suchthat \mv t_{\bar S}=\mv s\right\}.\]
%For any tuple $\mv t_{\bar S}$,
%define the {\em $\mv t_{\bar S}$-section} of $R$ as
%\[ R[\mv t_{\bar S}] = \pi_{\bar A\setminus \bar S}(R \lJoin \set{\mv t_{\bar S}}).
%\]

\section{Analysis of Algorithm~\ref{ALG:TRIANGLE-LEAP}}
\label{sec:triangle-leap}

It turns out that the run time of Algorithm~\ref{ALG:TRIANGLE-LEAP} is dominated by the time spent in computing 
the set $L_C^{a,b}$ for every $a\in L_A$ and $b\in L_B^a$. 
We now analyze this time ``inside out." We first note that for a given 
$a\in L_A$ and $b\in L_B^a$, the time it takes to compute $L_C^{a,b}$ is at 
most
\begin{equation}
 \min\left\{|\pi_C\sigma_{B=b}S|, |\pi_C\sigma_{A=a}T|\right\}\notag\\
\le \sqrt{|\pi_C\sigma_{B=b}S|\cdot|\pi_C\sigma_{A=a}T|},
\label{eq:leap-sum}
\end{equation}
where the inequality follows from~\eqref{eq:min-ub}. We now go one level up 
and sum the run time over all $b\in L_B^a$ (but for the same fixed $a\in L_A$ 
as above). This leads to the time taken to compute $L_C^{a,b}$ over all 
$b\in L_B^a$ to be upper bounded by
\begin{align*}
 &\sqrt{|\pi_C\sigma_{A=a}T|} \cdot\sum_{b\in L_B^a} \sqrt{|\pi_C\sigma_{B=b}S|}\\
\le &\sqrt{|\pi_C\sigma_{A=a}T|} \cdot\sqrt{|L_B^a|}\cdot \sqrt{\sum_{b\in L_B^a} |\pi_C\sigma_{B=b}S|}\\
\le &\sqrt{|\pi_C\sigma_{A=a}T|} \cdot\sqrt{|L_B^a|}\cdot \sqrt{\sum_{b\in \pi_BS} |\pi_C\sigma_{B=b}S|}\\
= &\sqrt{|\pi_C\sigma_{A=a}T|} \cdot\sqrt{|L_B^a|}\cdot \sqrt{|S|}\\
= &\sqrt{|S|}\cdot \sqrt{|\pi_C\sigma_{A=a}T|} \cdot\sqrt{|L_B^a|}\\
\le &\sqrt{|S|}\cdot \sqrt{|\pi_C\sigma_{A=a}T|} \cdot\sqrt{|\pi_B\sigma_{A=a}R|},
\end{align*}
where the first inequality follows from~\eqref{eq:cs} (where we have $L=L_B^a$ and the vectors are the all $1$'s vector and the vector $(\sqrt{|\pi_C\sigma_{B=b}S|})_{b\in L_B^a}$). The second inequality follows from the fact that $L_B^a\subseteq \pi_BS$. %The second equality uses the fact that $|S|=N$. 
 The final inequality follows from the fact that $L_B^a\subseteq \pi_B\sigma_{A=a}R$.

To complete the runtime analysis, we need to sum up the last expression above for every $a\in L_A$. However, note that this sum is exactly the expression~\eqref{eq:two-sum}, which we have already seen is $N^{3/2}$, as desired.

For completeness, we show how the analysis of 
Algorithm~\ref{ALG:TRIANGLE-LEAP} above follows directly from
Lemma~\ref{lmm:Qdecomp}. 
In this case we use Lemma~\ref{lmm:Qdecomp} with $I=I_2$, which implies 
\[L=R\Join \pi_BS\Join \pi_AT=\{(a,b)|a\in L_A, b\in L_B^a\},\]
where $L_A$ and $L_B^a$ is as defined in Algorithm~\ref{ALG:TRIANGLE-LEAP}.
Note that in this case $\calE_J=\{\{B,C\},\{A,C\}\}$. Thus, we have that the LHS in~\eqref{eqn:keyinequality} is the same as
\begin{align*}
&\sum_{a\in L_A}\sum_{b\in L_B^a} \sqrt{|S\lJoin (a,b)|}\cdot\sqrt{T\lJoin (a,b)|}\\
=&\sum_{a\in L_A}\sum_{b\in L_B^a} \sqrt{|\pi_C\sigma_{B=b}S|}\cdot\sqrt{|\pi_C\sigma_{A=a}T|}.
\end{align*}
Note that the last expression is the same as the one in~\eqref{eq:leap-sum}. Lemma~\ref{lmm:Qdecomp} argues that the above is at most $N^{3/2}$, which is exactly what we proved above.

\section{Technical Tools}

%\subsection{The generalized H\"older's inequality}

The following form of H\"older's inequality (also historically attributed to
Jensen) can be found in any standard texts on inequalities.
The reader is referred to the classic book ``Inequalities'' by
Hardy, Littlewood, and P\'olya \cite{MR89d:26016} (Theorem 22 on page 29).
%and the beautifully written monograph by Steele \cite{MR2062704}.

\begin{lmm}[H\"older inequality]
\label{lem:holder}
Let $m,n$ be positive integers. Let $y_1,\dots,y_n$ be non-negative
real numbers such that $y_1+\cdots+y_n\geq 1$.  Let $a_{ij} \geq 0$ be
non-negative real numbers, for $i\in [m]$ and $j\in [n]$.  With the
convention $0^0 = 0$, we have:
\begin{equation}
\sum_{i=1}^m\prod_{j=1}^n a_{ij}^{y_j}
\leq
\prod_{j=1}^n\left(\sum_{i=1}^m a_{ij}\right)^{y_j}.
\label{ineq:Holder}
\end{equation}
\end{lmm}

\section{Entropy and Alternate Derivations}
\label{app:entropy}

\subsection{Entropy and Shearer's inequality}

For basic background knowledge on information theory and the entropy function 
in particular, the reader is referred to Thomas and Cover \cite{MR2239987}.
We are necessarily brief in this section.

Let $X$ be a discrete random variable taking on values from a domain $\calX$
with probability mass function $p_X$.
The (binary) {\em entropy function} $H[X]$ is a measure of the degree of 
uncertainty associated with $X$, and is defined by
\[ H[X] := - \sum_{x\in \calX} p_X(x) \log_2 p_X(x). \]
We can replace $X$ by a tuple $\mv X = (X_1,\dots,X_n)$ of random variables 
and define the {\em joint entropy} in exactly the same way. The only 
difference in the formula above is the 
replacement of $p_X$ by the joint probability mass function.

Let $X,Y$ be two discrete random variables on domains $\calX, \calY$,
respectively. The {\em conditional entropy function} of
$X$ given $Y$ measures the degree of uncertainty about $X$ given that we knew
$Y$:
%\begin{multline*}
\[
 H[X \suchthat Y] := 
 - \sum_{\substack{x\in\calX\\y\in\calY}} \Pr[X=x,Y=y] \log_2 \Pr[X=x
\suchthat Y=y].
\]
%\end{multline*}
Again, we extend the above definition in the natural way when $X$ and $Y$ are
replaced by tuples of random variables.
Many simple relations regarding entropy and conditional entropy can be derived
from first principle, and they often have very intuitive interpretation. For
example, the inequality 
\[ H[X \suchthat Y, Z] \leq H[X \suchthat Y] \]
can be intuitively ``explained'' by thinking that knowing less (only $Y$ as
opposed to knowing both $Y$ and $Z$) leads to more uncertainty about $X$.
Similarly, the following formula can easily be derived from first principles.

\begin{equation}
H[X_1,\dots,X_n] = \sum_{j=1}^n H[H_j \suchthat X_1,\dots,X_{j-1}].
\label{eqn:cond-entropy}
\end{equation}

The above basic observation can be used to prove Shearer's inequality below.
Our proof here follows the conditional entropy approach from Radhakrishnan
\cite{radhakrishnan20036}. The version of Shearer's lemma below is slightly
more direct than the version used in \cite{AGM08, grohebounds}, leading to a 
shorter proof of \agm's inequality in Section~\ref{subsec:AGMproof}.

\blmm[Shearer \cite{MR859293}]
Let $X_1,\dots,X_n$ be $n$ random variables. For each subset $F \subseteq [n]$,
let $\mv X_F = (X_i)_{i\in F}$; and, let $\mv X = X_{[n]}$.
Let $\calH = (\calV = [n], \calE)$ be a hypergraph, and
$\mv x = (x_F)_{F\in \calE}$ be any fractional edge cover of the hypergraph.
Then, the following inequality holds
\begin{equation}
H[\mv X] \leq \sum_{F\in \cal E} x_F \cdot H[\mv X_F]
\label{eqn:shearer}
\end{equation}
\elmm
\bp
For every $F\in\calE$ and $j\in F$, we have
\[ 
   H[X_j \suchthat X_i, i<j]
   \leq
   H[X_j \suchthat X_i, i<j, i \in F] 
\]
because the entropy on left hand side is conditioned on a (perhaps non-strict)
superset of the variables conditioned on the right hand side.
Additionally, because the vector $\mv x = (x_F)_{F\in\calE}$ is a fractional
edge cover, for every $j \in [n]$ we have 
$\sum_{F\in\calE, j\in F} x_F \geq 1.$
It follows that, for every $j \in [n]$,
\[ H[X_j \suchthat X_i, i<j] \leq
   \sum_{F\in\calE, j\in F} x_F \cdot H[X_j \suchthat X_i, i<j, i \in F]. 
\]
From this inequality and formula \eqref{eqn:cond-entropy}, we obtain
\begin{eqnarray*}
H[\mv X]   
&=& \sum_{j=1}^n  H[X_j \suchthat X_i, i<j]\\
&\leq& \sum_{j=1}^n\sum_{F\in\calE, j\in F} x_F \cdot H[X_j \suchthat X_i, i<j, i \in F]\\
&=& \sum_{F\in\calE} x_F \cdot \sum_{j \in F} H[X_j \suchthat X_i, i<j, i \in F]\\
&=& \sum_{F\in\calE} x_F \cdot H[\mv X_F].
\end{eqnarray*}
\ep

\subsection{\agm's proof based on Shearer's entropy inequality and a 
join-project plan}
\label{subsec:AGMproof}

\agm's inequality was shown in \cite{AGM08, GM06} as follows. 
A similar approach was used in \cite{MR1639767} to prove an essentially
equivalent inequality regarding the number of copies of a hypergraph in 
another.
Let $\mv X = (X_v)_{v\in\calV}$ denote a uniformly chosen random tuple from 
the output of the query $Q$. 
Then, $H[\mv X] = \log_2 |Q|$. Note that each $X_v, v\in \calV$ is a
random variable. For each $F\in\calE$, let $\mv X_F = (X_v)_{v\in F}$. Then, the
random variables $\mv X_F$ takes on values in $R_F$.
Because the uniform distribution has the maximum entropy,
we have $H[\mv X_F] \leq \log_2 |R_F|$, for all $F\in\calE$.

Let $\mv x$ denote any fractional edge cover for the hypergraph
$\calH=(\calV,\calE)$ of the query $Q$, then from Shearer's
inequality~\eqref{eqn:shearer} we obtain
\[
\log_2|Q| = H[\mv X]
\leq \sum_{F\in\calE} x_F \cdot H[\mv X_F]
\leq \sum_{F\in\calE} x_F \cdot \log_2 |R_F|,
\]
which is exactly \agm's inequality~\eqref{eqn:AGM}.

\bprop
For any query $Q$, there is a join-project plan 
with runtime $O(|Q|^2 \cdot 2^{\rho^*(Q,\calD)} \cdot N)$ for evaluating $Q$, 
where $N$ is the input size.
\eprop
\bp
We define the join-project plan recursively.
Let $A_1,\dots,A_n$ be the attributes
and $R_1,\dots,R_m$ be the relations of $Q$.
Let $\bar B_{n-1} = (A_1,\dots,A_{n-1})$. 
We first recursively compute the join 
$P = \ \Join_{F\in \calE} \pi_{\bar B_{n-1}}(R_F).$
Then, we output 
\[ Q = (\cdots ((P \Join \pi_{A_n}(R_1)) \Join \pi_{A_n}(R_2) ) 
        \Join \cdots \Join \pi_{A_n}(R_m)). 
\]
The base case is when $n=1$ which is the simple $m$ set intersection problem.
It is easy to see that $\rho^*(P, \calD) \leq \rho^*(Q, \calD)$, because 
any fractional edge cover for $Q$ is also a fractional edge cover for $P$.
Hence, all the intermediate results in computing $Q$ from $P$ has sizes at most
$|P| \cdot N \leq 2^{\rho^*(P, \calD)} N \leq 2^{\rho^*(Q,\calD)} \cdot N$.
From there, the claimed run-time follows.
\ep

\section{A more formal description of \glvv results}
\label{app:glvv}

From examples in Section~\ref{subsec:conjunctive}, we can describe 
\glvv's strategy for obtaining
size bounds for general conjunctive queries with simple functional 
dependencies. Let $C$ be such query. The general idea is to construct
a natural join query $Q$ such that $|C| \leq |Q|$ and thus we can simply apply
\agm bound on $Q$.
\bi
 \item The first step is to turn the conjunctive query $C$ into 
 $C_1 = $ chase$(C)$. 
 
This step can be done in $O(|C|^4)$-time \cite{Maier:1979:TID:320107.320115,
 DBLP:journals/tods/AhoBU79, DBLP:journals/jacm/BeeriV84,
 DBLP:books/aw/AbiteboulHV95}. This is the step in the spirit of Example
 \ref{ex:chase}.
 \item Next, let $C_2$ be obtained from $C_1$ by replacing each repeated 
 relation symbol by a fresh relation symbol. Thus, in $C_2$ there is no 
 duplicate relation. (In the join algorithm itself, all we have to do is to 
 have them point to the same underlying index.)
 \item The next step is in the spirit of Example \ref{ex:FD}. Recall that a
 simple functional dependency (FD) is of the form $R[i] \to R[j]$, which means
 the $j$th attribute of $R$ is functionally determined by the $i$th attribute.
 From the set of given simple FDs, we can obtain a set of FDs in terms of the
 variables in the query $C_2$. For example, if $R[3] \to R[1]$ is a given FD 
 and $R(XYZW)$ occurs in the body of the query $C_2$, then we obtain a FD of 
 the form $Z \to X$. Now, WLOG assume that $\vars(C_2) = \{X_1,\dots,X_n\}$.
 We repeatedly apply the following steps for each (variable) FD of the form
 $X_i \to X_j$:
  \bi
    \item For every atom $R$ in which $X_i$ appears but $X_j$ does not, add
    $X_j$ to $R$. (This is a fictitious relation -- and in the join algorithm 
    we do not need to physically add a new field to the data; rather, we keep a
    pointer to where $X_j$ can be found when needed.)
    \item For every FD of the form $X_k \to X_i$, we add $X_k \to X_j$ as a new
    FD.
    \item Remove the FD $X_i \to X_j$.
  \ei
It is easy to see that performing the above in a canonical order will 
terminate in about $O(n^2)$ time, and the new query $C_3$
(with some fictitious relations ``blown up'' with more attributes) is exactly
equivalent to the the old query $C_2$.
\item Next, we can remove repeated variables from relations by keeping only
tuples that match the repetition patterns. For example, if there is an atom
$R(XXY)$ in the query $C_3$, then we only have to keep tuples $(t_1,t_2,t_3)$ in
$R$ for which $t_1=t_2$. After this step, we can replace $R$ by a relation
$R'(XY)$. This step is in the spirit of Example \ref{ex:repeated}.
We call the resulting query $C_4$.
\item Finally, we turn $C_4$ into $Q$ by ``projecting out'' all the attributes
not in the head atom as was shown in Example \ref{ex:projection}.
Since $Q$ is now a join query, \agm bounds applies.
\ei
What is remarkable is that \glvv showed that the bound obtained this way is 
essentially tight up to a factor which is data-independent.

Note again that our description above is essentially what was obtained from
\glvv, without the coloring number machinery.
Furthermore, if $C$ was a full conjunctive query, then we don't have to project
away any attribute and thus any worst-case join algorithm described in the
previous section can be applied which is still worst-case optimal!

\section{Proof of Lemma \ref{LEM:BAD:INSTANCE}}
\label{sec:bad-instance}

\bp[Proof of Lemma \ref{LEM:BAD:INSTANCE}]
In the instances below the domain of any attribute is
$$\calD = \{0,1,\dots,(N-1)/(n-1)\},$$
where we  ignore the integrality issue for the sake of clarify.
For $i\in [n]$, let $R_i$ denote the set of {\em all} tuples
in $\calD^{[n]-\{i\}}$ each of which has at most one non-zero value.
It follows that, for all $i \in [n]$,
\[ |R_i| = (n-1)[(N-1)/(n-1)+1] - (n-2) = N, \]
and that
\begin{eqnarray*}
 |\Join_{i=1}^n R_i| &=& n[(N-1)/(n-1)+1]-(n-1)\\
 &=& N+(N-1)/(n-1).
\end{eqnarray*}

(We remark that the instance above was specialized to $n=3$ in Section~\ref{sec:triangle}.)

A relation $R$ on attribute set $\bar A \subseteq [n]$ is called ``simple"
if $R$ is the set of {\em all} tuples in $\calD^{\bar A}$ each of which has at
most one non-zero value. Then, we observe the following properties.

\bi
 \item[(a)] The input relations $R_i$ are simple.
 \item[(b)] An arbitrary projection of a simple relation is simple.
 \item[(c)] Let $S$ and $T$ be any two simple relations on attribute sets
  $\bar A_S$ and $\bar A_T$, respectively.  If $\bar A_S$ is contained in 
  $\bar A_T$ or vice versa, then $S \Join T$ is simple.
  If neither $\bar A_S$  nor $\bar A_T$ is contained in the other,
  then $|S\Join T| \geq (1+(N-1)/(n-1))^2 = \Omega(N^2/n^2)$.
\ei

For an arbitrary join-project plan starting from the simple relations
$R_i$,
we eventually must join two relations whose attribute sets are not
contained
in one another, which right then requires $\Omega(N^2/n^2)$ run time.
\ep

\end{document}